\documentclass[preprint,12pt]{elsarticle}

\usepackage{graphicx}
\usepackage{amssymb}
\usepackage{lineno}

\usepackage{lineno,hyperref}
\usepackage{subfig} % For the subfloat command
\usepackage{siunitx} % siunitx typesets physical units in a consistent manner
\usepackage{booktabs} % booktabs provides professional formatting commands for tables
\usepackage{amsmath} % amsmath provides extra maths symbols
\usepackage{graphicx, amssymb, wasysym,epsf, graphicx, outlines, enumitem, tikz}
\usepackage{chemformula}
\usepackage{textcomp} % textcomp provides extra text symbols (like a degrees celsius symbol)

 \usepackage[final]{changes}

\journal{Journal of Colloid and Interface Science}

\begin{document}

\begin{frontmatter}


\title{Rheology of protein-stabilised emulsion gels envisioned as composite networks.\\ 1 - Comparison of pure droplet gels and protein gels}



\author[Ad:Unilever,Ad:UoE]{Marion Roullet\corref{cor1}}
\ead{marion.roullet@espci.org}
\author[Ad:UoE]{Paul S. Clegg}
\ead{paul.clegg@ed.ac.uk}
\author[Ad:Unilever]{William J. Frith}
\ead{bill.frith@unilever.com}

\address[Ad:Unilever]{Unilever R\&D Colworth, Sharnbrook, Bedford, MK44 1LQ, UK}
\address[Ad:UoE]{School of Physics and Astronomy, University of Edinburgh, Peter Guthrie Tait Road, Edinburgh, EH9 3FD, UK}
\cortext[cor1]{Current address: BioTeam/ECPM-ICPEES, UMR CNRS 7515, Universit\'{e} de Strasbourg, 
25 rue Becquerel, 67087 Strasbourg Cedex 2, France}

\begin{abstract}
	%Word limit: 200 words

%Because of their ability to strongly adsorb at oil/water interfaces and to stabilise oil droplets by steric and electrostatic repulsion, water-soluble proteins are widely used as efficient emulsifiers. At low pH, proteins can aggregate and also form gels, either of protein molecules in solution or of protein-covered droplets. Despite the industrial relevance of these gels, their rheological properties ...

%The soft materials formed from protein-stabilised emulsions, like yogurt, are referred to as emulsion gels, however, this designation is not precise enough to reflect the variety of composition of these materials. It is thus important to take into account the ratio between adsorbed and suspended proteins in solution before gelation, as during emulsification not all the proteins in solution adsorb at the interface. If the ratio is low, i.e. most of the protein is suspended, the system can be seen as a matrix of protein gel with the oil droplets acting as fillers, as in polymeric materials. At a higher ratio the emulsion gel is more of a composite formed of both a protein and a droplet network.
%Our objective is to study in detail the protein-stabilised emulsion gels considering the full range of their composition. A first step is to characterise separately the gelation of purified suspensions of protein-stabilised droplets, and of suspensions of pure proteins. These components are then combined, resulting in emulsion gels of well-characterised compositions, thus allowing a rigorous approach to these systems.

	\subsection*{Hypothesis}
Protein-stabilised emulsion gels can be studied in the theoretical framework of colloidal gels, because both protein assemblies and droplets may be considered as soft colloids. These particles differ in their nature, size and softness, and these differences may have an influence on the rheological properties of the gels they form.
	
	\subsection*{Experiments}
Pure gels made of milk proteins (sodium caseinate), or of sub-micron protein-stabilised droplets, were prepared by slow acidification of suspensions at various concentrations. Their microstructure was characterised, their viscoelasticity, both in the linear and non-linear regime, and their frequency dependence were measured, and the behaviour of the two types of gels was compared.
	\subsection*{Findings}
Protein gels and droplet gels were found to have broadly similar microstructure and rheological properties when compared at fixed volume fraction, a parameter derived from the study of the viscosity of the suspensions formed by proteins and by droplets. The viscoelasticity displayed a power law behaviour in concentration, as did the storage modulus in frequency. % which seems to indicate the particle like behaviour persists after gelation for both systems.
Additionally, strain hardening was  found to occur at low concentration. These behaviours differed slightly between protein gels and droplet gels, showing that some specific properties of the primary colloidal particles play a role in the development of the rheological properties of the gels.

%Whilst ungelled samples show clear soft particle like behaviour, the gelled systems show a more nuanced behaviour, beyond the volume fraction scaling.
%These results can be used to understand the contributions of each component in the rheological properties of emulsion gels containing both proteins and protein-stabilised droplets.	

\end{abstract}

\begin{keyword}
Colloidal gel \sep Rheology \sep Emulsion \sep Sodium caseinate \sep Viscoelasticity \sep Protein-stabilized droplet \sep Microstructure
%Need american spelling, up to 10 keywords


\end{keyword}

%%Mandatory graphical abstract: space in template or need a new one?

\end{frontmatter}

\section*{Graphical abstract}
\begin{figure}[htbp]
	\begin{center}
		\includegraphics[width=13cm]{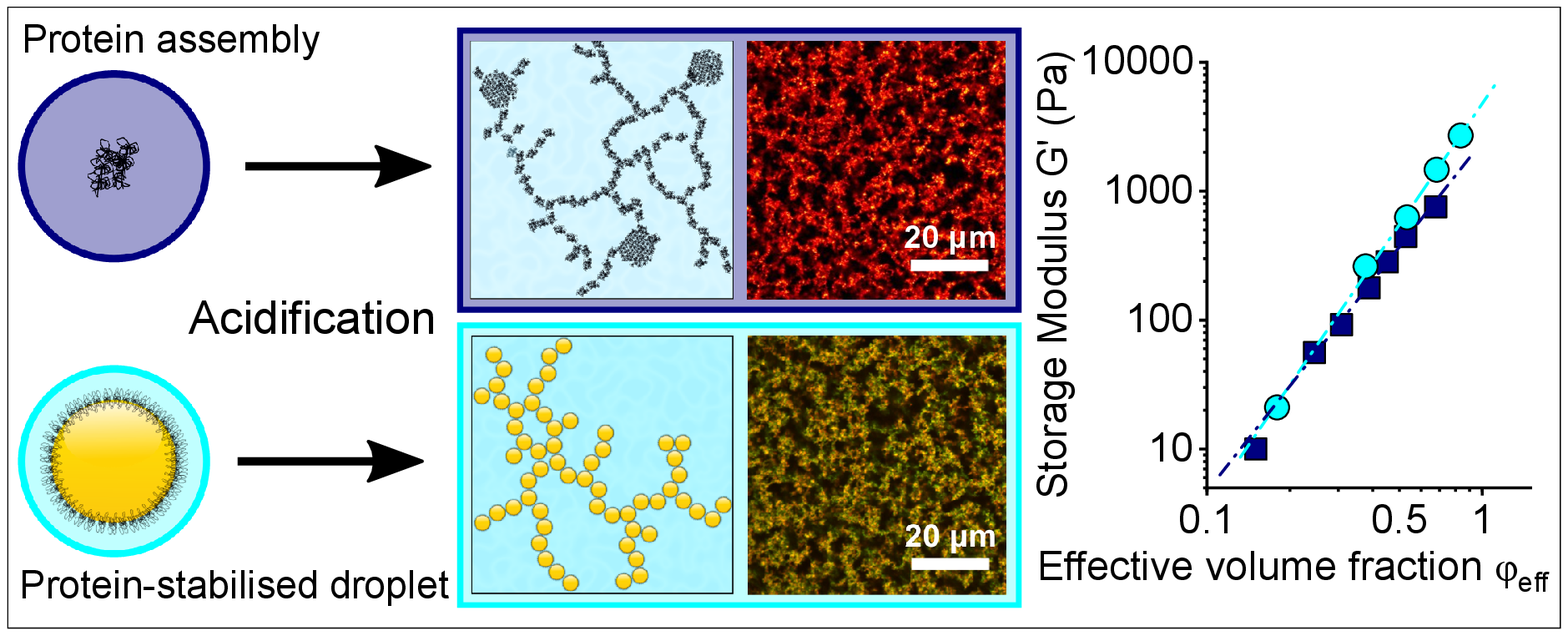}
	\end{center}
	\label{Fig:GraphAbstract}
\end{figure}

%\linenumbers
\begin{figure}[h]
		\includegraphics[width=2cm]{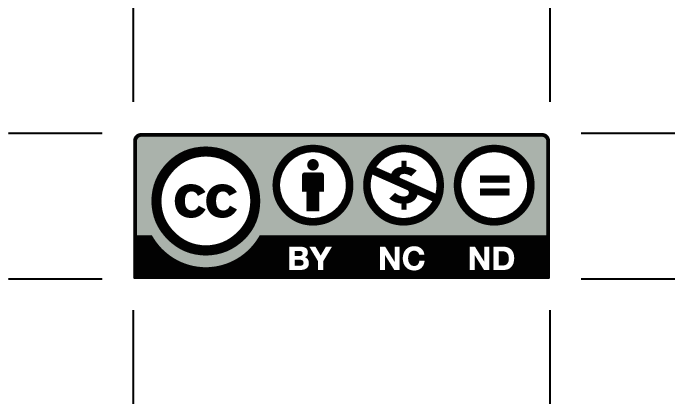}
\end{figure}\textcopyright 2020. This manuscript version is made available under the \href{http://creativecommons.org/licenses/by-nc-nd/4.0/}{CC-BY-NC-ND 4.0 license}

\section{Introduction}

Emulsion and protein gels form the basis of many food products, such as yoghurt, soft cheese or tofu, where the flocculation of a vegetable or animal milk leads to the formation of a soft solid via aggregation of proteins and fat droplets. This process has been used for millenia in traditional cooking, but a deep understanding of the mechanisms of the physical transformation occurring in these systems only came in recent decades with the study of colloidal gels \cite{dickinson:2011, kasapis:2009:2,delgado:2016}. While much effort has been spent in correlating the structure formation and the gel properties with the interparticle interactions \cite{zaccarelli:2007}, there is yet to be a full understanding of food-based colloidal gels, both in terms of fundamental science and of specific applications.

%To be written again: gels of interest here typically fractal with fractal dimension found to be ... in literature. Known that power law variation of G' with volume fraction, and power law can be related to fractal dimension, easily if DLCA and less easily in more general case (Wu and Morbidelli)
\added{The gels of interest here typically exhibit a fractal microstructure that can be described by a  fractal dimension $D_f$  \cite{vanvliet:2000}. This fractal microstructure affects their mechanical properties \cite{walstra:1991, bremer:1990}. The storage modulus $G'$ of the gels also typically shows a power-law variation with the volume fraction $\phi$ \cite{krall:1998,wu:2001}:} 
\begin{equation}
G'\sim \phi^A
\label{Eq:StorModGel}
\end{equation}
\added{It has been shown that the exponent $A$ can be related to the fractal dimension of the gels, with the relationship depending on the gelation regime. For gels formed via diffusion limited cluster aggregation, generally at low volume fractions, it was found that $A=(3+D_b)/(3-D_f)$, where  $D_b$ is defined as the bond (or backbone) dimension of the network \cite{zaccarelli:2007, delgado:2016}. At higher volume fractions, the links between clusters are weaker and  $A=1/(3-D_f)$ \cite{shih:1990}. A general model, as suggested by Wu and Morbidelli, is $A=(1+(2+D_b)(1-\epsilon))/(3-D_f)$, where $\epsilon\in\left[0;1\right]$ depends on the type of regime \cite{wu:2001}.}
\deleted{The gels of interest here typically form via diffusion limited cluster aggregation (\textit{DLCA})  %\cite{zaccarelli:2007, delgado:2016}
	, which result in a fractal microstructure %{vanvliet:2000}
	 that can be described by a  fractal dimension $D_f$. This fractal microstructure affects their mechanical properties% \cite{walstra:1991, bremer:1990}
	 . The storage modulus $G'$ of the gels can be related to the fractal dimension $D_f$ and to the volume fraction $\phi$ by the following %\cite{krall:1998}
 :$G'\sim \phi^\frac{3+D_b}{3-D_f}$}

\deleted{Where  $D_b$ is defined as the bond (or backbone) dimension of the network. $D_b$ usually has a value close to $1$, and, in the case of a DLCA gel, the stress-bearing backbone is relatively linear, yielding a bond dimension of $D_b\approx\num{1.1}$ \cite{delgado:2016}. The storage modulus of a DLCA gel thus varies with the volume fraction $\phi$ of particles following a power law of exponent $\left(3+D_b\right)/\left(3-D_f\right)$, which has been experimentally observed for colloidal gels \cite{krall:1998, gibaud:2013, aime:2018}.}
This theoretical framework for colloidal gels can be applied not only to model attractive hard spheres, but also to protein and emulsion gels, and in particular to casein systems \cite{vanvliet:2000, dickinson:1996, chen:1999a}.

Caseins are the most common proteins in cow's milk. They have attracted considerable attention for the last 40 years, mainly because of their widespread use as food ingredients in numerous commercial products (processed cheese, ice-cream, coffee whiteners, cream liqueur, etc). In this study, sodium caseinate, which is derived from the caseins in milk, was used both as gelling agent and as emulsifier. 

Sodium caseinate, when suspended in water, forms naturally-occurring aggregates, that are thought to be elongated with a length around $\SI{20}{\nano\metre}$ \cite{farrer:1999,lucey:2000,huppertz:2017}. The surface of these aggregates is charged negatively at neutral pH, and electrostatic repulsion is an important condition for their stability \cite{dickinson:2006}. When such suspensions are acidified, the decrease in electrostatic repulsion causes the aggregation of proteins that, if slow and rather homogeneous, leads to the formation of a gel \cite{phillips:1994:9, chen:1999a, mellema:2002, takeuchi:2008, braga:2006}.

Previous work using confocal microscopy has highlighted their fractal structure, which was found to be dependent on the pH, ageing time and addition of other components \cite{pugnaloni:2005,mellema:2000, walstra:1991,leocmach:2015, mahmoudi:2015,moschakis:2010}. A power-law dependence of the viscoelasticity on concentration of acid casein gels, using both native casein and sodium caseinate, has been observed in previous studies \cite{bremer:1990, mellema:2002}, which was attributed to their fractal nature \cite{chen:1999a, vanvliet:2000}. In addition, the frequency dependence of the gels has been characterised \cite{chen:1999a, bremer:1990, ruis:2007} . Finally, the brittle fracture of casein gels has also been studied from a fundamental perspective \cite{saint-michel:2017, leocmach:2014}.

Besides gel formation, sodium caseinate is widely used to form oil-in-water emulsions  \cite{dickinson:1988:1,dickinson:1992,dickinson:2001}. Typically, during emulsification, the proteins do not completely adsorb at the interface, leaving a residual fraction of protein suspended in the continuous phase after emulsification \cite{srinivasan:1996,srinivasan:1999}, and the resulting emulsion is thus a mixture of droplets and of un-adsorbed proteins \cite{roullet:2019a}. Consequently, the distinction is made here between protein-stabilised emulsions, and purified droplet suspensions, from which the fraction of un-adsorbed proteins was removed.

As with caseinate gels, the acidification of sodium caseinate-stabilised emulsions, and of purified droplet suspensions, leads to the formation of fractal gels called \textit{emulsion gels} \cite{dickinson:2006}, and of pure droplet gels respectively. For pure caseinate-stabilised droplet gels, the nature of the interactions at play during gelation is the same as for caseinate gels, as the droplets become attractive at the isoelectric point of the protein.

Emulsion gels have been studied in the past and compared to protein gels \cite{chen:1999a,dickinson:1996, dickinson:2006}, and they have been shown to present a similar fractal structure \cite{dickinson:2003}. However, \deleted{a thorough comparison of these systems is still lacking. In addition,} the properties of pure droplet gels have not been clearly investigated, \added{in that the systems studied have invariably contained both droplets and free protein. This has made it extremely difficult to draw a consistent comparison of protein assemblies and protein-stabilised droplets as gel-forming particles.} Investigating the pure components - droplets and proteins - would enable a consistent comparison of their behaviours and understanding of their mixtures\added{, that would be relevant from both a fundamental and a technological point of view}.

%Paragraph to emphasis importance of study: theory (2 colloidal gels with supposedly same attraction but different particle nature and softness) and technology (replace one by the other in food system, common issue in formulation, low fat vs cheap, but how to do that rigorously?)

To this end, the present study investigates the similarities and differences between caseinate gels and pure caseinate-stabilised droplet gels over a wide range of concentrations. It focuses more specifically on the microstructure and on key rheological features of these systems, namely the linear and non-linear viscoelasticity and the frequency dependence, as well as on their variations with the concentration in colloidal species.

This comparison between caseinate gels and droplet gels draws on the results of a previous study of the viscosity of the suspensions that are used to prepare these gels \cite{roullet:2019a}. It was shown that both droplets and protein assemblies can be studied in the framework developed for soft colloidal particles \cite{vlassopoulos:2014}. Consequently, their concentrations can be scaled by the effective volume fraction $\phi_{eff}$\added{, which can reach high values due to the possible compression, interpenetration and deformation of soft colloids} \cite{conley:2017,winkler:2014}. It is demonstrated in the present study how the same concentration scaling can be used to understand the behaviour of both sodium caseinate and droplet gels.

\section{Materials \& Methods}

	\subsection{Preparation of protein and droplet suspensions}
Suspensions of pure sodium caseinate (hydrodynamic radius $\SI{11}{\nano\meter}$) and of pure sodium caseinate-stabilised droplets (hydrodynamic radius $\SI{110}{\nano\meter}$) were prepared at a range of concentrations, as described previously \cite{roullet:2019a}. The procedure is given in detail in Section 1.1 of the supplementary information for completeness.
%Composition of samples? 
These suspensions were then used as sols for the preparation of acid-induced gels.

In the following, concentrations of both the protein and droplet samples are given in terms of the effective volume fraction $\phi_{eff}$. This was determined from intrinsic viscosity measurements on dilute samples as detailed in Ref. \cite{roullet:2019a}. As such the weight concentration is related to $\phi_{eff}$ by a simple factor, which was found to be $\SI[separate-uncertainty=true]{2.2(1)}{\milli\liter\per\gram}$ and $\SI[separate-uncertainty=true]{8.5(2)}{\milli\liter\per\gram}$ for the droplet and protein suspensions respectively. The use of this parameter is discussed in detail in Ref. \cite{roullet:2019a}.

	\subsection{Preparation of protein and droplet gels}
	The decrease in pH required for the gelation of the suspensions of sodium caseinate and of pure sodium-caseinate stabilised droplets to occur was achieved by the slow hydrolysis of glucono $\delta$-lactone (Roquette), as detailed in Section 1.2 of the supplementary material.

	%Although this ratio was found empirically, it can be related to the ratio glucono $\delta$-lactone:protein if the layer of adsorbed protein is assumed to represent $\SI{13}{\percent}$ of the weight of the droplets, as will be detailed in Chapter~\ref{Chap:Charac}. In this case, it is found that $\frac{glucono~\delta-lactone}{protein}=\num{0.58}$ for droplet suspensions. The discrepancy with the value for protein suspensions probably originates from the differences in configuration between the two colloidal systems, as part of the protein may be buried inside the self-forming sodium caseinate aggregates, while the protein is probably unfolded at the surface of the oil droplet, thus requiring more acidifier to reach the same pH.
	%0.075 g gdl/ g drop (because 0.0105 g gdl for a 7g sample of 20\% droplets): if assume that $[prot]_{drop}=0.1$ then 0.75g gdl / g prot for droplets (but protein not buried inside aggregates).
	
%	For mixtures of proteins and droplets, the amount of glucono $\delta$-lactone was calculated appropriately for the protein and droplet contents of each sample. The final pH was kept between $\num{4.5}$ and $\num{5.0}$ as in this range of pH, caseinate is at its isoelectric point and thus forms strong gels \cite{chen:1999a}.

	\subsection{Laser scanning confocal microscopy}

The gels were imaged using laser scanning confocal microscopy, here a setup based on an LSM 780  microscope on inverted Axio observer (Zeiss). It has to be noted that the resolution of confocal microscopy (limited to $\approx\SI{200}{\nano\meter}$ by light diffraction) does not allow imaging of the single protein aggregates, or single droplets. Instead, the lengthscale accessible by this imaging technique corresponds to the structure over a few colloidal particles, and is thus suitable for the description of colloidal aggregation and gelation.

\subsubsection{Protocol for imaging of gels}
\label{Sec:MicroProtocol}

Rhodamine B (Sigma Aldrich)  and Bodipy 493/503 (Molecular Probes) were added to the samples of protein and droplet suspensions, that were then mixed with glucono $\delta$-lactone, as detailed in Section 1.3 of the supplementary material.

\subsection{Image analysis}
The image analysis of 2D micrographs was performed using the image processing software \textsf{ImageJ} \cite{imagej}.

The Fast Fourier Transform (FFT) analysis was applied to the image after Hanning windowing. The image in the Fourier space was then radially averaged to obtain the spectrum $I(q)$. The wave vector $q$ represents a spatial frequency, it is a function of the distance from the centre in the Fourier space and of the image size, and is expressed in $\SI{}{\per\micro\meter}$.

\begin{figure}[htbp]
	\begin{center}
		\includegraphics[width=0.8\textwidth]{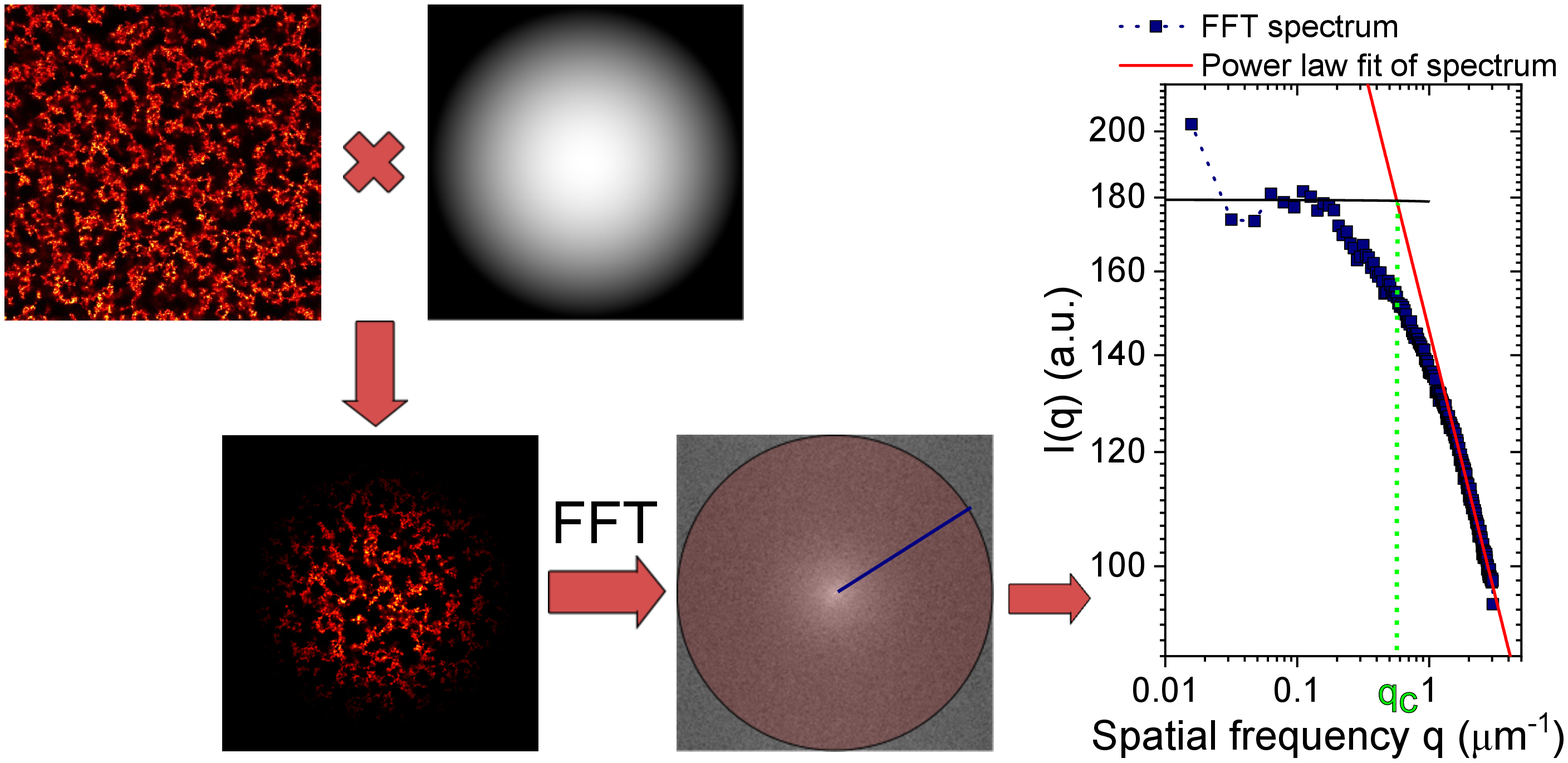}
	\end{center}
	\caption{Protocol for the analysis of a micrograph. \\
	The image is multiplied by a Hanning window of the same size before the Fast Fourier Transform  is calculated. The spectrum $I(q)$ is obtained using the plugin \textsf{Radial Profile} to perform a radial average of the Fourier transform.\\
    The decrease of the spectrum $I(q)$ is then fitted by a power law, linear in double logarithmic scale. Its intersection with the plateau defines the critical spatial frequency $q_c$.}
	\label{Fig:FFTProt}
\end{figure}

The variations of $I(q)$ can be described by several parameters. The position of the shoulder $q_c$ was chosen in this study as critical wave vector, because it can be estimated in a reproducible way by fitting the power law decrease of the peak, as opposed to the top of the peak that is slightly flattened. The determination of the position of the shoulder $q_c$ is illustrated in Figure~\ref{Fig:FFTProt}. This value was then used to estimate the critical lengthscale of the network in the real space $L_C=2\pi/q_c$.

	\subsection{Rheological measurements}
Oscillatory rheology measurements were performed using a stress-controlled MCR 502 rheometer (Anton Paar) and a Couette geometry ($\SI{17}{\milli\metre}$ diameter profiled bob and cup CC17-P6, inner diameter $\SI{16.66}{\milli\meter}$, outer diameter $\SI{18.08}{\milli\meter}$ yielding a $\SI{0.71}{\milli\meter}$ tool gap, gap length $\SI{25}{\milli\meter}$). To avoid slip at the wall during shearing, profiled bob and cup (serration width $\SI{1.5}{\milli\meter}$, serration depth $\SI{0.5}{\milli\meter}$) were selected as measurement tools.
The temperature was set by a Peltier cell at $\SI{35}{\celsius}$ during the entire measurement sequence. \added{The operating temperature was chosen to ensure that the gelation occurs in the time scale of thousands of seconds for all the samples studied here.} To prevent evaporation, a thin layer of silicon oil of low viscosity ($10$ cSt) was deposited on the surface of the sample.

The measurements were started immediately after mixing of the sample with glucono $\delta$- lactone and subsequent loading in the instrument. As represented in Figure~\ref{Fig:RheoMeasSeq},the measurements consisted of 4 steps, detailed in Section 1.4 of the supplementary information, \added{for which the strain was chosen to stay in the linear viscoelastic region at steps 1, 2 and 4}.

\begin{figure}[htb]
	\begin{center}
		\includegraphics[width=0.65\textwidth]{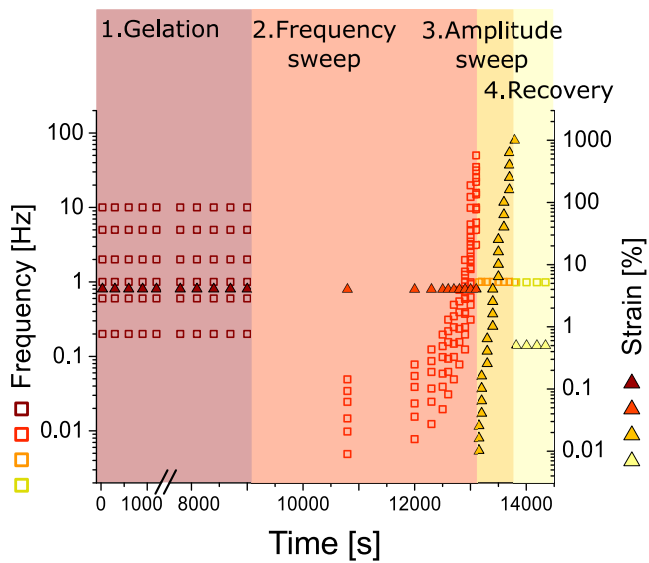}
	\end{center}
	\caption{Illustration of the measuring sequence for the oscillatory rheometry of the emulsion gels, detailed in the Methods section. Frequency (open squares) and strain amplitude (filled triangles) of the oscillatory shear vary with time in the 4 steps of the measurement. The multiwave mode was activated at steps 1 and 2, leading so several signal frequencies were used simultaneously.  At the time $t=\SI{0}{\second}$, the glucono $\delta$-lactone was added to the sols.}
	\label{Fig:RheoMeasSeq}
\end{figure}

For each sample, $3$ measurements of the same batch of sample were performed and the values of each data point were averaged. 

\section{Results \& Discussion}

%Structure for rheology: presentation of results (and fits if needed), discussion in terms of food collloid (comparison with literature, consequences for gels), discussion in terms of rheology (models, comparison with colloidal gels and/or simu, explanation)

%Comparison, how we proceed
The comparison of pure caseinate-stabilised droplet gels and caseinate gels was performed by studying each type of system over a wide range of concentrations, scaled by the effective volume fractions $\phi_{eff}$. This extensive characterisation of each type of system ensured that the similarities and differences observed derived from the intrinsic differences in size, structure and softness between caseinate assemblies and caseinate-stabilised droplets. This precaution distinguishes the present study from previous comparisons of emulsion gels and protein gels \cite{rosa:2006, chen:1999a} and is the key to the progress made here.

\subsection{Microstructure of gels: colloidal species and volume fraction}
\label{Sec:MicroProtDrop}

\begin{figure}[htbp]
	\begin{center}
		\includegraphics[width=0.80\textwidth]{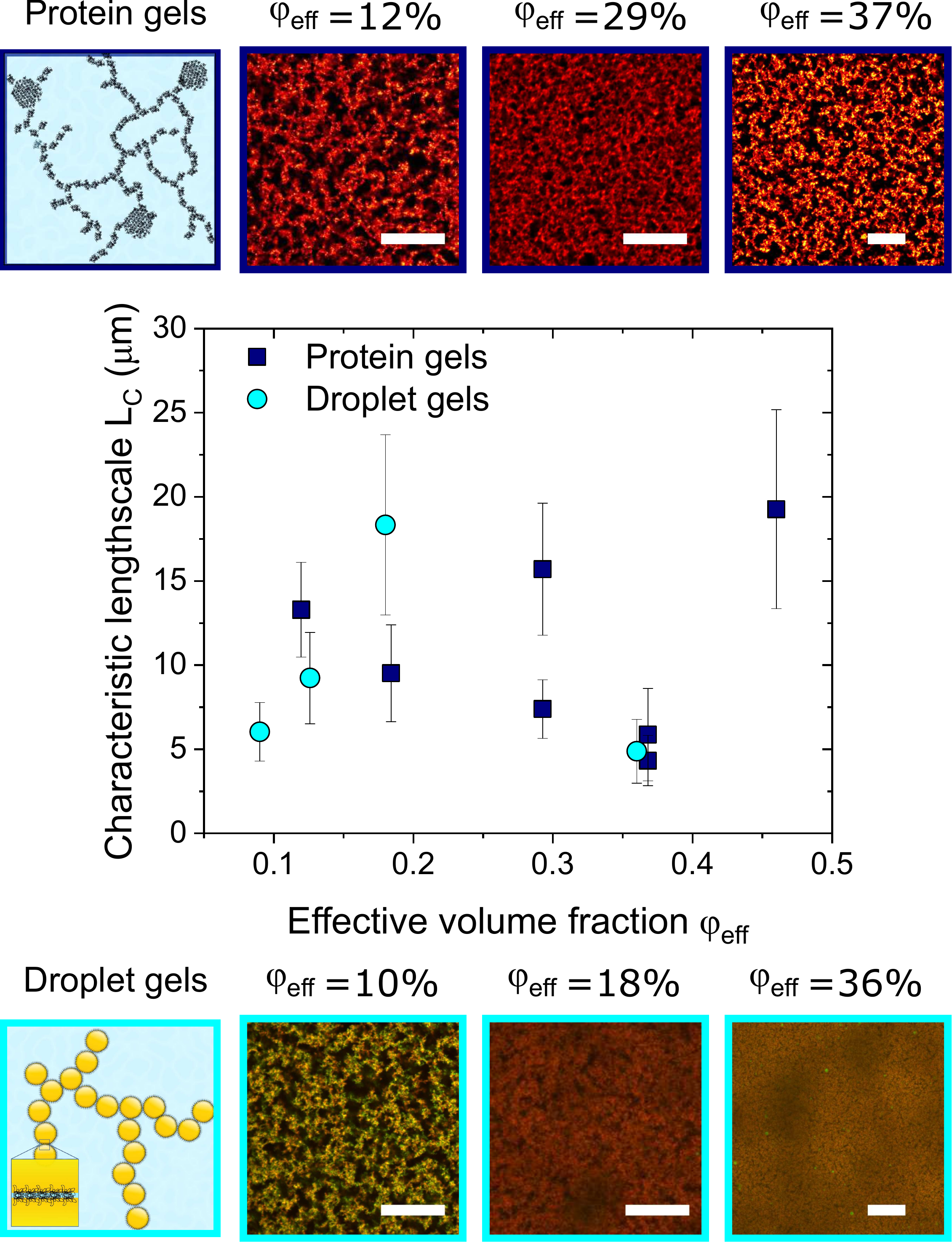}
	\end{center}
	\caption{Micrographs ($\SI{100}{\micro\meter} \times \SI{100}{\micro\meter}$) of aged acid-induced gels formed from suspensions of: (top) sodium caseinate, and of (bottom) caseinate-stabilised droplets, at different volume fractions $\phi_{eff}$. The scale bars are $\SI{30}{\micro\meter}$ long. \added{The inset in the cartoon representing the droplet gels shows the interactions between the caseinate adsorbed at the oil-water interface.} The graph presents the characteristic lengthscale $L_C$ of the gels, as a function of the volume fraction $\phi_{eff}$, for caseinate gels (squares, navy blue) and caseinate-stabilised droplet gels (circles, cyan). \\ For each point, $L_C$ was obtained by performing a FFT of one micrograph and extracting the position of the peak in the spectrum $I(q)$, as described in Figure~\ref{Fig:FFTProt}. The inaccuracy of this determination is indicated by the error bar. Where two points are presented for one concentration, they correspond to different micrographs of similar samples. }
	\label{Fig:ConfocalGels}
\end{figure}

Confocal microscopy is a commonly used technique to observe the structure of colloidal gels at the micron scale \cite{dinsmore:2001,stradner:2004,prasad:2007} that makes possible the comparison of this structure for gels of different composition and volume fraction. Here, gels that were prepared by acidifying suspensions of either sodium caseinate or pure caseinate-stabilised droplets, at different concentrations, are imaged and compared. %As described in Section~\ref{Sec:MicroProtocol}, the gels were imaged a long time after gelation, and considered to be in an approximated equilibrium. The image analysis was performed as described in Figure~\ref{Fig:FFTProt}. 
The micrographs of caseinate and droplet gels, together with their characteristic lengthscale $L_C$ are presented in Figure~\ref{Fig:ConfocalGels}.

As can be seen, the micrographs are similar for protein and droplet gels, especially at lower volume fraction. Indeed, in both cases, the fractal structure typical of colloidal gels is present, with interconnected networks of particle aggregates (in colour) and water-filled pores (in black). At high concentrations, these networks are denser in particles, with the pores of the droplet gels appearing to be smaller than for the protein gels.

%Notably, for droplet gels, overlap between proteins labeled by Rhodamine B (in red) and oil droplets (in green). This is because lengthscale is 10 times larger than droplet size.

In addition, the characteristic lengthscales $L_C$ are of the same order of magnitude for the two components, and their values range between $\SI{5}{\micro\meter}$ and $\SI{20}{\micro\meter}$ for all the gels presented here. \replaced{The variation of $L_C$ as a function of the volume fraction cannot be interpreted quantitatively because of the significant noise in the data. This is partially related to the fact that the features picked up by the Fast Fourier Transform are probably a combination of the size of the aggregates and the size of the pores.}{Because the features picked up by the Fast Fourier Transform are probably a combination of the size of the aggregates and the size of the pores, the variation of $L_C$ as a function of the volume fraction cannot be interpreted in detail.}

%This is reflected by the discrepancy of $L_C$ for two different micrographs of the same protein gel at  $\phi_{eff}=\SI{30}{\percent}$. 
%It is however expected that the mesh size decreases with increasing volume fraction, as has been observed numerically for colloidal gels at a lower range of volume fractions ($\phi=\SI{2.5}{\percent}$ to $\phi=\SI{10}{\percent}$) \cite{colombo:2014a}. In the framework of the fractal theory, this decrease in lengthscale with the volume fraction can be understood as a decrease in the cluster size required to reach a volume-spanning network.

Thus, although the individual droplets are one order of magnitude larger than the individual protein assemblies, the gels formed by these two types of colloidal particles present a very similar fractal structure at a given effective volume fraction $\phi_{eff}$. It would also be interesting to perform a more thorough investigation of the dependence of the characteristic length scale $L_C$ of the network on the effective volume fraction $\phi_{eff}$, by using higher quality confocal micrographs and a more precise image analysis technique\added{, for example texture analysis microscopy \cite{gao:2014}}. %It is also interesting to note that the effective volume fraction scaling, derived from the study of suspensions, does not break down for the gels, but works instead over a larger lengthscale.

%The fractal structure of the gels and the presence of interconnected networks of pores and of particle clusters make their quantification difficult. Here, the FFT was chosen to determine a characteristic length scale $L_C$ of the networks, that appeared to be in the same order of magnitude for all the micrographs presented. In addition, it may be possible to extract other features from the images, such as the size distribution of the pores, the fractal dimension and the connectivity of the network. This information could possibly bring a deeper understanding of the structure of protein and droplet gels, but would require higher quality confocal micrographs, as any pre-treatment of the image may induce a bias. The fractal dimension for example, despite being clearly defined in theory, has been shown to poorly describe casein gels \cite{pugnaloni:2005, mellema:2000}.

\subsection{Rheological study of droplet gels and protein gels}
In order to investigate further the comparison between protein gels and droplet gels, it is interesting to characterise their rheological behaviour. As detailed in Figure~\ref{Fig:RheoMeasSeq}, the viscoelastic moduli, $G'$ and $G''$, are first compared at fixed frequency, strain and time after gelation for gels of different compositions. Then the dependences of $G'$ and $G''$ on the frequency are presented. Finally, the non-linear viscoelasticity of the gels is considered.

\subsubsection{Linear viscoelasticity of gels}

\begin{figure}[htbp]
	\centering
	\includegraphics[width=\textwidth]{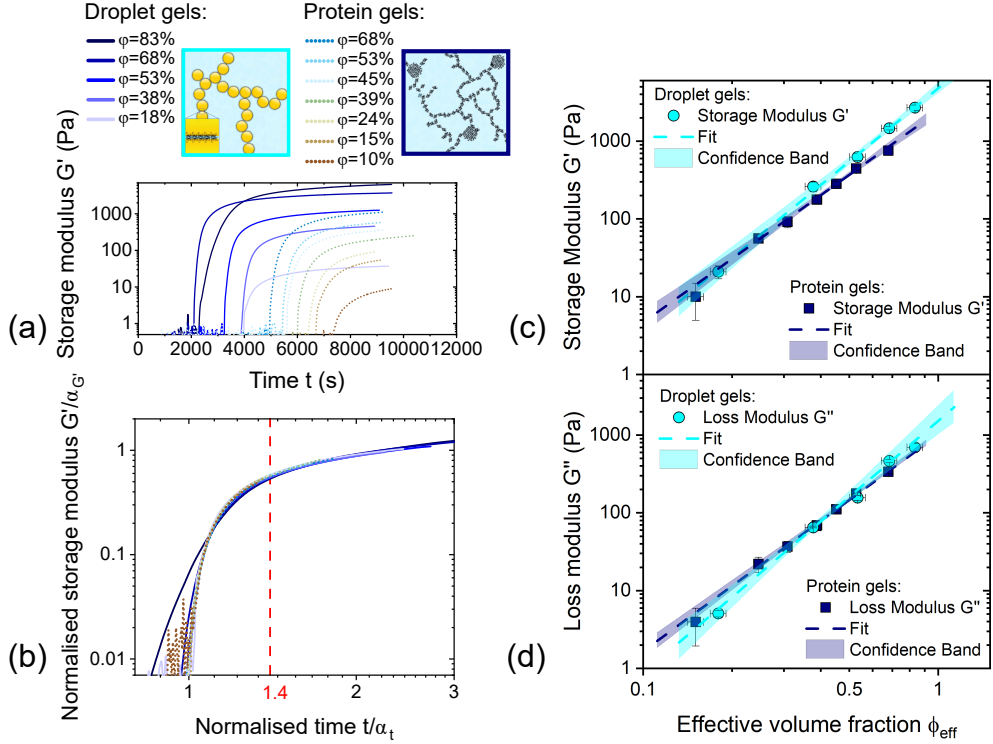}
	\caption{Left panel: (a) Storage moduli G' upon formation of droplet gels and protein gels, (b) Master curve for the formation of the colloidal gels, the horizontal ($\alpha_t$) and vertical ($\alpha_{G'}$) shift factors are presented in Figure S3. Right panel: Storage ($G'$, (c)) and loss ($G''$, (d)) moduli at $\SI{1}{\hertz}$  of protein-stabilised droplet gels (circles, cyan) and of protein gels (squares, navy blue) at $t/\alpha_t=1.4$ as functions of the effective volume fraction of the gel $\phi_{eff}$. A fit (Equation~\ref{Eq:PowerLaw}) of each type of system was performed and the model (parameters listed in Table~\ref{Tab:PowerLaw}) as well as the $\SI{95}{\percent}$ confidence band are displayed on each graph. \\The horizontal error bars arise from error propagation upon calculation of the volume fraction, as detailed in Section 3 of the supplementary material, while the vertical error bars arise from the uncertainties in determining the shift factors $\alpha_{G'}$ and $\alpha_t$.}
	\label{Fig:GelsPowerLaws}
\end{figure}

The gelation of sodium caseinate and sodium caseinate-stabilised droplets \added{is presented in Figure~\ref{Fig:GelsPowerLaws}(a). As with previous studies on colloidal gels, these systems do not reach an equilibrium state, but go through rearrangements of their network upon ageing \cite{cipelletti:2000,colombo:2013}. To compare the viscoelasticity of the gels at similar ageing state, it is possible to superimpose the gelation curves by using horizontal and vertical shifts in logarithmic scale \cite{ruis:2007,meunier:1999,calvet:2004}, as can be seen in Figure~\ref{Fig:GelsPowerLaws}(b). The horizontal and vertical shift factors $\alpha_t$ and $\alpha_{G'}$, and the protocol used to determine them, can be found in Figure S3 and Section 2 of the supplementary material.}

\deleted{has been studied previously %\cite{chen:1999a}
	, so the results obtained here are only presented in Section 2 of the supplementary material. As with previous studies on colloidal gels, these systems do not reach an equilibrium state, but go through rearrangement of their network upon ageing %\cite{cipelletti:2000,colombo:2013}
	. To compare the viscoelasticity of the gels at similar ageing time, the storage and loss moduli $G'$ and $G''$ were measured at $\SI{2500}{\second}$ after gelation. For both gels of protein-stabilised droplets and of proteins, the influence of the effective volume fraction was studied, and the behaviours were compared.}

%\added{The two types of gels present a remarkably similar gelation behaviour, except for the most concentrated droplet gel, at $\phi_{eff,drop}=0.83$. While this sample displays a zero-shear viscosity before acidification, it is very viscous, and it is thus possible }

%Error for gels: estimated from fit procedure (semi-manual: slight change of data selection for the fit, and look at variation of parameter) From error shift factors (both vertical and horizontal), determine what is the range of G' we can get
\added{The visoelastic behaviour of gels was arbitrarily compared at $t/\alpha_t=1.4$. This value was chosen because it is the highest reached by all the gels studied, even those with a very slow gelation. Because the kinetics that determine $\alpha_t$ remain the same post gelation, the rise in elastic modulus $G'$ with $t/\alpha_t$ is similar for all samples. Thus using $G'$ at constant $t/\alpha_t$ is appropriate for comparison of different concentrations.} The elastic modulus $G'$ and the loss modulus $G''$ of the two types of gels \added{at $t/\alpha_t=1.4$}, measured at $\SI{1}{\hertz}$, are presented in Figure~\ref{Fig:GelsPowerLaws} as functions of their effective volume fraction $\phi_{eff}$.
\replaced{In addition, the phase angle $\delta=arctan(G''/G')$, indicating the viscoelastic character of the gels, is found to be significantly different for each sort of gels, with $\delta_{prot}$ varying between $\SI{21}{\degree}$ and $\SI{24}{\degree}$, and $\delta_{drop}$ between $\SI{13}{\degree}$ and $\SI{17}{\degree}$.}{In addition, the phase angle $\delta=arctan(G''/G')$, indicating the viscoelastic character of the gels, is found to be relatively constant for each sort of gels, with $\delta_{prot}=\SI{21}{\degree}$ and $\delta_{drop}=\SI{13}{\degree}$.}  The higher phase angle found for protein gels indicates that their behaviour is slightly shifted towards the viscous materials on the spectrum of viscoelastic behaviour.

As can be seen in Figure~\ref{Fig:GelsPowerLaws}, sodium caseinate and sodium caseinate-stabilised droplets form gels of very similar viscoelasticity when scaled by the volume fraction. More precisely, the storage and loss moduli of droplet gels are slightly higher, at a given volume fraction, than those of protein gels. The similarity of the viscoelasticity of the two types of gels can be related to their similar microstructure, as observed in Figure~\ref{Fig:ConfocalGels}.

Our result differs significantly from a previous study on caseinate-stabilised emulsion gels \cite{chen:1999a}. Although the concentrations chosen for the comparison were arbitrary in Ref.~\cite{chen:1999a}, it was shown that emulsion gels had a similar modulus to a protein gel with a threefold increase in protein concentration, and the authors thus concluded that emulsions form gels with a higher viscoelasticity than protein gels. It is thought that this discrepancy arises mostly from the choice of parameter to describe the composition of these systems. Indeed, the storage modulus can be presented as a function of either the protein concentration, or of the weight concentration of each colloidal particle, leading to large differences between protein gels and droplet gels,  but in opposite directions, as illustrated in Figure S4 of the supplementary material.   %Indeed, if the moduli were plotted as functions of the weight concentration of each colloidal particle, the two lines would be shifted horizontally, and the moduli of a protein gel would be almost two orders of magnitude higher than the moduli of a droplet gel at the same concentration. If the protein concentration was instead chosen to describe the droplet system, as proteins are estimated to represent only roughly $\SI{10}{\percent}$ of the weight of the droplets, then the droplet gels would appear to be much stiffer than the protein gel \cite{chen:1999a}. Regardless of its definition, 

More generally, the weight concentration is unlikely to be a relevant parameter to compare gels made of colloidal particles of a very different nature, such as caseinate assemblies and droplets - the former being water-swollen and soft, while the latter are filled with oil and more rigid. The same is true for the use of the protein concentration, as shown in Ref.~\cite{chen:1999a}. Instead, we argue here that a more appropriate scaling to use for comparing protein gels and droplet gels is the volume fraction, despite its definition being non-trivial for complex colloidal particles \cite{roullet:2019a}.% here the volume fraction is approximated by the effective volume fraction $\phi_{eff}$. 

Consequently, we find that there is little difference between the two types of gels, provided that the comparison is drawn between samples at the same effective volume fraction $\phi_{eff}$. Furthermore, the variation of the viscoelasticity with the volume fraction for the protein gels and the droplet gels can be quantified by using a fit to a power law, as discussed below.

\paragraph{Power-law increase with volume fraction}

As can be seen in Figure~\ref{Fig:GelsPowerLaws}, the variations of both storage $G'$ and loss $G''$ moduli as functions of effective volume fraction can be described as a power law for the two types of gels: 
\begin{equation}
G(\phi_{eff}) = G_{0,\phi} \times \phi_{eff}^{\alpha}
\label{Eq:PowerLaw}
\end{equation} 
Where the pre-factor of the power-law $G_{0,\phi}$ and the exponent $\alpha$ are two parameters to be determined. The values found by fitting $G'(\phi_{eff})$ and $G''(\phi_{eff})$ with Equation~\ref{Eq:PowerLaw} are summarised in Table~\ref{Tab:PowerLaw}.

\begin{table}[hbt]
	\centering
	\caption{Parameters for Equation~\ref{Eq:PowerLaw} to fit viscoelasticity of gels at $\SI{1}{\hertz}$ displayed in Figure~\ref{Fig:GelsPowerLaws}}
	\begin{tabular}{|l|cc|cc|}
		\hline
& \multicolumn{2}{c|}{Storage modulus $G'$} & \multicolumn{2}{c|}{Loss modulus $G''$} \\
\cline{2-5}
Gel type	&  $G'_{0,\phi}$  & $\alpha$  & $G''_{0,\phi}$  & $\alpha$ \\
\hline
%Droplet gels & \replaced{\SI{4.79}{\kilo\pascal}}{\SI{6.93}{\kilo\pascal}} & \replaced{\num[separate-uncertainty=true]{3.1(1)}}{\num[separate-uncertainty=true]{3.24(7)}} & \replaced{\SI{1.52}{\kilo\pascal}}{\SI{1.37}{\kilo\pascal}} & \replaced{\num[separate-uncertainty=true]{3.2(1)}}{\num[separate-uncertainty=true]{3.11(9)}} \\
Droplet gels & \SI[separate-uncertainty=true]{4.78(22)}{\kilo\pascal} & \num[separate-uncertainty=true]{3.1(1)} & \SI[separate-uncertainty=true]{1.52(21)}{\kilo\pascal} & \num[separate-uncertainty=true]{3.2(1)} \\
%Protein gels & \replaced{\SI{2.43}{\kilo\pascal}}{\SI[separate-uncertainty=true]{3.1(4)}{\kilo\pascal}} & \replaced{\num[separate-uncertainty=true]{2.7(1)}}{\num[separate-uncertainty=true]{3.05(11)}} & \replaced{\SI{1.01}{\kilo\pascal}}{\SI{1.11}{\kilo\pascal}} & \replaced{\num[separate-uncertainty=true]{2.8(1)}}{\num[separate-uncertainty=true]{2.91(11)}} \\
Protein gels & \SI[separate-uncertainty=true]{2.42(19)}{\kilo\pascal} & \num[separate-uncertainty=true]{2.7(1)} & \SI[separate-uncertainty=true]{1.01(3)}{\kilo\pascal} & \num[separate-uncertainty=true]{2.8(1)} \\
\hline
	\end{tabular}
	\label{Tab:PowerLaw}
\end{table}

This power-law dependence of the viscoelasticity of sodium caseinate gels is in good correspondence with previous studies on casein gels \cite{walstra:1991, bremer:1990, mellema:2002, ruis:2007, mahmoudi:2015}. \replaced{The value of the exponent for sodium caseinate varies significantly with temperature, as it was found that $\alpha = \num{2.57}$ at $\SI{30}{\celsius}$ and $\alpha = \num{3.73}$ at $\SI{50}{\celsius}$ \cite{ruis:2007}. The value found here for gels formed at $\SI{35}{\celsius}$ is thus in good agreement with these results}{The value of the exponent for sodium caseinate varies significantly with the experimental conditions, like temperature and ageing time, as it was found to vary from $\alpha = \num{2.57}$ \cite{ruis:2007} to $\alpha = \num{4.6}$ \cite{bremer:1990} , a precise comparison of the value found here with the literature may thus not be relevant}. In addition, no data is available on the rheological properties of acid-induced droplet gels.

%This result corroborates the observations made for the micro-structure, and confirms that the volume fraction scaling is a relevant approximation to study colloidal gels. Indeed, if the gels were compared using their weight concentration, the discrepancy in volume fraction could be mistaken for an intrinsic difference between protein and emulsion gels. Despite its definition being non-trivial for complex colloidal particles, the volume fraction is thus an essential parameter that reveals the similarity between protein gels and droplet gels.

\deleted{It has to be noted that t}The power law dependence of the elastic modulus $G'$ is a feature of fractal colloidal gels, as previously observed experimentally and numerically \cite{vanvliet:2000,krall:1998, gibaud:2013, aime:2018,deoliveira:2019,buscall:1987}\added{, that can be related to the fractal dimension $D_f$. However, the large range of volume fractions for the gels presented here makes such analysis impractical, in the absence of additional characterisation of these networks.}. \deleted{Consequently, the results displayed in Table~\ref{Tab:PowerLaw} can be examined in light of the theoretical framework developed for such systems.}
%The variation of the viscoelasticity of each type of gels with the volume fraction can be discussed more in details by using the theoretical framework of colloidal gels.

\deleted{The viscoelasticity of a colloidal gel is closely related to its fractal structure %\cite{delgado:2016,aime:2018}
	. In particular, the storage modulus $G'$ is a direct consequence of the presence of a stress-bearing network within the material. The dependence of $G'$ on the volume fraction of particles $\phi$ can thus be linked to the way the elementary particles are arranged, described by the fractal dimension $D_f$.}

% In this framework, the exponent $\alpha$ of the power law presented in Figure~\ref{Fig:GelsPowerLaws} is expressed \cite{delgado:2016, krall:1998}:
%\begin{equation}
%\alpha=\frac{3+D_b}{3-D_f}
%\label{Eq:FracDim}
%\end{equation}
%Where the bond dimension $D_b$ reflects the fractal dimension of the gel backbone. For gels formed by attractive particles in the dilute regime, the gelation most likely occurs by diffusion-limited cluster aggregation (DLCA), and in that case the stress-bearing backbone is relatively linear, yielding a bond dimension of $D_B\approx\num{1.1}$.

\deleted{In this framework, Equation~\ref{Eq:StorModGel} can be applied to the protein and droplet gels, and the fractal dimension is found to be $D_f=3-(3+D_B)/\alpha\approx\num{1.7}$. This value is in good correspondence with previous results for DLCA particle gels %\cite{asnaghi:1992, krall:1998}
	, which indicates that the formation of protein gels and droplet gels follow this mechanism. The detailed nature and energy profile of interparticle interaction involved in sodium caseinate aggregation is not known, but this result seems to correspond with strong Van der Waals attractions when the proteins are at their isoelectric point.}

\deleted{However, it has to be noted that the dependence of $G'$ on $\phi$ is only a proxy for determining the fractal dimension, and is greatly limited by the wide range of concentrations explored here. Indeed, if a DLCA regime is expected for gels formed from semi-dilute suspensions ($\phi_{eff}\approx\SI{10}{\percent}$), the mechanisms are likely to be different for gels at $\phi_{eff}\approx\SI{70}{\percent}$, in which particle crowding will play an important part in the structure formation, and thus the fractal dimension may differ.}% Despite the lack of unifying theory for the viscoelasticity of colloidal gels at different concentrations, the power-law variation of the linear elastic modulus $G'$ with the volume fraction $\phi$ has also been observed for gels formed of rubber particles \cite{deoliveira:2019}.

%About value of G'_0        
\deleted{In addition to the exponent $\alpha$, the pre-factor of the power law $G'_{0,\phi}$ can also be discussed in the theoretical framework of colloidal gels. Droplet gels appear to form slightly stiffer gels when scaled by the volume fraction, as $G'_{0,\phi,drop}\approx\num{2}\times G'_{0,\phi, prot}$. An explanation for this difference could be the discrepancy in size of the particles, or a difference in bond topology between protein assemblies and protein-stabilised droplets. Indeed, a difference in bond strength may appear if a flattened patch of adsorbed proteins is formed upon contact between bonded droplets. In this hypothetical case, the bond strength between colloidal particles would be larger for droplet gels than for protein gels.}

%There is also a model for fractal colloidal gels in \cite{shih:1990}: cf notes
%Use Mellema categorisation of scaling? \cite{mellema:2002},
%cf Vanv vliet 2000 , also comparison with particle gels in \cite{dickinson:2013}
%If the structure formed is influenced by Brownian motion during gel formation (check van Vliet \cite{walstra:1991}), can normalise the moduli by thermal energy (cf work on glass by petekidis et al) \cite{koumakis:2012}.
%Normalise by density of contacts: divide by volume of particle (assume it is spherical) $4/3 \Pi R^3$. Divide or multiply? Multiplication gives energy.

The study of the gel moduli as a function of their composition, described both by the nature of the elementary particles and by their volume fraction, thus offers some information on the mechanical properties of caseinate gels and caseinate-stabilised droplet gels. The behaviour of the two types of gels is very similar and reminiscent of those of more model colloidal gels. In addition to this static view of protein and emulsion gels, it is important to compare their dynamic properties.

\subsubsection{Frequency dependence of gels}

%Intro:
%Why is frequency dependence important? Related to stability in time of final product (also related to syneresis? Look for literature)
%So good to characterise this feature and compare the behaviours of the two types of gel, 

%How does it compare with Divoux's results?

The moduli of the newly formed gels were then measured over a wide range of frequency. This measurement of the frequency dependence makes it possible to probe the dynamics of the gels. Because these \replaced{exhibit a solid behaviour in the linear viscoelastic range}{are solids}, this aspect is limited to fluctuations within the gel network, for example rearrangement of the particle bonds, relaxation of the stress bearing strands, or motion of non-stress bearing strands like dangling chains. 

\paragraph{Comparison between protein gel and droplet gel}

\begin{figure}[htb]
	\begin{center}
		\includegraphics[width=0.9\textwidth]{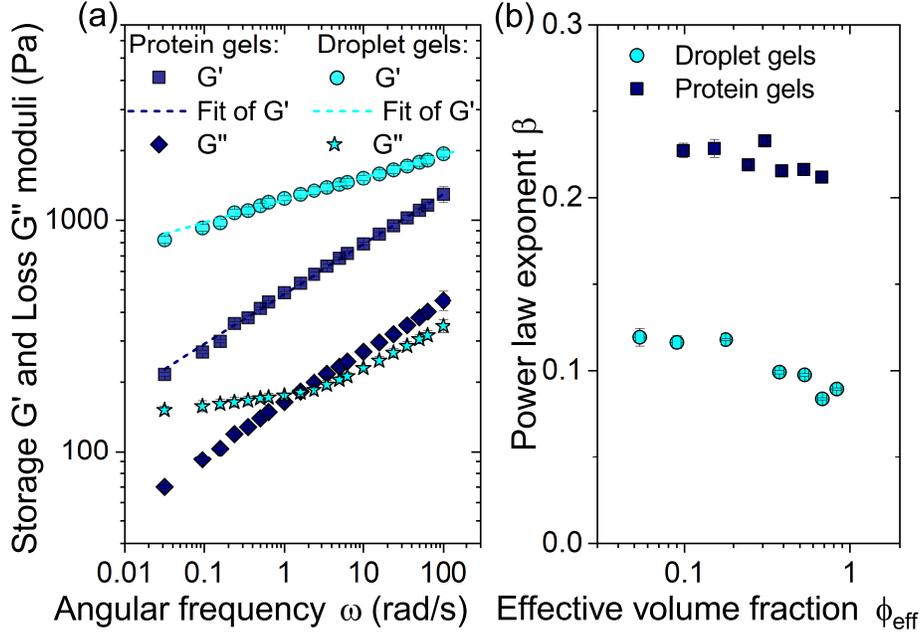}
	\end{center}
	\caption{(a) Comparison of the frequency dependence for protein gels (sodium caseinate: $\phi_{eff}=\SI{53}{\percent}$, in navy blue) and droplet gels (caseinate-stabilised oil droplets: $\phi_{eff}=\SI{53}{\percent}$, in cyan). Storage modulus $G'$ and loss modulus $G''$ are represented as functions of the angular frequency $\omega$. $G'$ was fitted with a power law for both types of samples, and the fitting parameters can be found in Table~\ref{Tab:FreqPowerLaw}.\\
	(b) Comparison of frequency dependence for protein gels (squares, navy blue) and protein-stabilised droplets (circles, cyan): power-law exponent $\beta$, obtained by fitting $G'=f(\omega)$ with Equation~\ref{Fig:CompFreq}, as a function of the effective volume fraction $\phi_{eff}$.}
	\label{Fig:CompFreq}
\end{figure}

In order to compare similar gels of proteins and of protein-stabilised droplets, gels of equal volume fraction ($\phi_{eff}=\num{0.53}$) are displayed in Figure~\ref{Fig:CompFreq} (a).

%General behaviour
Both the protein gel and the droplet gel exhibit an increase of their viscoelasticity with the angular frequency $\omega$, in agreement with previous studies on colloidal gels \cite{bremer:1990,ruis:2007,chen:1999a,bouzid:2018c}. The storage modulus $G'$ increases moderately for the two types of gels, while the loss modulus $G''$ also rises with $\omega$, but with a slightly different behaviour for protein gels and droplet gels. The increase of $G''$ is at odds with the frequency dependence of dilute colloidal gels, for which a decrease of $G''$ was observed \cite{aime:2018}, but is in good correspondence with the computed linear viscoelasticity of a similar system \cite{bouzid:2018c}. This behaviour may indicate the presence of a relaxation process that is visible in the frequency range covered at low concentration, but which moves to much lower frequencies at higher concentrations, and so becomes invisible.

%Differences between protein and droplet gels
In addition, as can be seen in Figure~\ref{Fig:CompFreq} (a), the viscoelastic response of the two types of gels differ slightly. Indeed, the protein gel displays a higher dependence on frequency than the droplet gel, as both storage and loss moduli increase faster with the angular frequency than for the droplet gel. Another noticeable difference is the non-monotonic behaviour of the loss modulus $G''$ for droplets gels. This behaviour may be an indication of a relaxation of droplet networks, that would be absent for protein gels in this range of frequency, but an extended spectrum would be required to definitely identify a possible peak. 

%When decrease concentration in droplets (ie G'), we also decrease max frequency that gives good results (over cut-off frequency, randomly high values for G' and G'')

In order to quantify the difference in variation of the storage modulus $G'$ with the angular frequency $\omega$ for the two types of gels, its behaviour can be modelled by a power law \cite{johnson:2019,jabbari-farouji:2008}: 
\begin{equation}
G'=G'_{0,\omega}~\left(\frac{\omega}{\omega_{\beta}}\right)^\beta
\label{Eq:FreqPowerLaw}
\end{equation} 
Where $G'_{0,\omega}$ and $\beta$ are two empirical parameters to be determined, and $\omega_{\beta}=\SI{1.00}{\radian\per\second}$ is used for dimensional purposes. 

The frequency dependence of the protein gel and droplet gel of effective volume fraction $\phi_{eff}=\SI{53}{\percent}$ is thus fitted as displayed in Figure~\ref{Fig:CompFreq} (a), and the values of the empirical parameters for can be found in Table~\ref{Tab:FreqPowerLaw}. %Or some theory about them?

\begin{table}[hbt]
	\centering
	\caption{Frequency dependence of gels: parameters from using Equation~\ref{Eq:FreqPowerLaw} to fit the variation of the storage modulus $G'$ with the angular frequency $\omega$ of protein gel and droplet gel of effective volume fraction $\phi_{eff}=\SI{53}{\percent}$ displayed in Figure~\ref{Fig:CompFreq}.}
	\begin{tabular}{|l|cc|}
		\hline
		Gel type	&  $G'_{0,\omega}$  & $\beta$  \\
		\hline
		Protein gels & \SI{0.5}{\kilo\pascal} & \num[separate-uncertainty=true]{0.22}\\
		Droplet gels & \SI{1.2}{\kilo\pascal} & \num[separate-uncertainty=true]{0.10} \\
		\hline
	\end{tabular}
	\label{Tab:FreqPowerLaw}
\end{table}

The value of the exponent $\beta$ for caseinate gels is slightly higher than in previous studies. Indeed, for acid-induced casein gels at $\SI{30}{\celsius}$, $\beta$ was measured to be $0.15$ \cite{bremer:1990,ruis:2007,chen:1999a}. This discrepancy may arise from a difference of pH of the gels studied, a parameter which was shown to have a strong influence on the frequency dependence of such systems \cite{chen:1999a}.

No comparable data could be found for the frequency dependence of gels made of pure protein-stabilised droplets, but the comparison between protein gels and gels of mixtures of proteins and droplets was performed and appears to be system-dependent. On one hand, the exponent $\beta$ was found to be identical for acid-induced gels of caseinate emulsions and for caseinate gels, i.e. $0.15$ \cite{chen:1999a}. On the other hand, for heat-set gels and emulsion gels prepared with $\beta$-lactoglobulin, %the gelation route was identified as an important factor for the frequency dependence. Indeed, no difference between emulsion and protein gels was observed when an enzyme was used to induce gelation as both gels displayed very little dependence on frequency, while for heat set gels
the slope $\beta$ was found to be three times higher for protein gels than for emulsion gels \cite{dickinson:1996}. This discrepancy is believed to result from the nature of the bonds between particles in these two types of gels: heat-set gels form more transient bonds than acid-induced gels, making for more mobile structures.

\paragraph{Influence of the volume fraction on the frequency dependence}

This analysis of the frequency dependence can be extended to gels at all concentrations, as the curves display a similar power-law variation. These can be found in Figure S6 of the supplementary information.
The empirical model in Equation~\ref{Eq:FreqPowerLaw} was thus applied to the gels of droplets and proteins prepared at different volume fractions, and the resulting value of the power law exponent for all the gels is displayed in Figure~\ref{Fig:CompFreq} (b).    

%\begin{figure}[htb]
%	\begin{center}
%		\includegraphics[width=0.9\textwidth]{GrFSExponent}
%	\end{center}
%	\caption{Comparison of frequency dependence for protein gels (squares, navy blue) and protein-stabilised droplets (circles, cyan): power-law exponent $\beta$, obtained by fitting $G'=f(\omega)$ with Equation~\ref{Fig:CompFreq}, as a function of the effective volume fraction $\phi_{eff}$. }
%	\label{Fig:FreqExponent}
%\end{figure}

As can be observed, there is little influence of the volume fraction on the variation of the elasticity of the gels with the frequency. This indicates that over the range of concentrations studied, the dynamical behaviour of the gels is the same. By contrast, the viscosity of the suspensions increases dramatically over the same range of volume fraction,  as discussed in a previous study \cite{roullet:2019a}. The negligible variations of the frequency dependence of the gels seem to indicate that there is no change in regime due to the crowding of the colloidal particles in the solid state, and the gels formed by proteins and droplets suspensions are similar in that respect.

Consequently, the difference in dynamic behaviour between protein gels and droplet gels observed in Figure~\ref{Fig:CompFreq} (a) is consistent over the range of volume fractions explored here, with the exponent $\beta$ being larger for protein gels than for droplet gels. This seems to indicate that caseinate gels have more internal fluctuations than droplet gels regardless of their concentration \cite{jabbari-farouji:2008}.%, which may be a reflection of the more pronounced rearrangement of protein gels observed by rheo-imaging during gelation.

Furthermore, the non-monotonic behaviour of the loss modulus $G''$ with the frequency $\omega$  observed for the droplet gel in Figure~\ref{Fig:CompFreq} (a) is also consistent over the range of volume fractions, as can be found in Figure S5. This is better visualised by looking at the phase angle of the gels, as presented in Figure S6. In contrast, the phase angle of all the protein gels studied is constant with frequency. The physical mechanism underlying this behaviour is not known but it represents an additional significant difference in the frequency dependence of droplet gels compared to protein gels. %Or should we give a possible explanation? Or ask someone who may know? Difference is significant only if we can relate it to some origin, otherwise not very interesting.

%Further comparison with gels made with microgels? cf Minami 2019, Appel 2016 "Mechanics at the glass-to-gel transition of thermoresponsive microgel suspensions"

Finally, these results of the linear viscoelasticity of colloidal gels can be compared with another sort of arrested state of colloidal particles, such as glasses of soft colloids \added{like microgels \cite{minami:2019}}. For such systems, it was observed that at moderately high volume fraction, the glasses display a slow increase in elastic modulus $G'$ with the frequency, associated with some mobility of the particles in an entropic glass. By contrast, at higher volume fraction, the particles are completely jammed and $G'$ is constant over the range of frequency explored \cite{pellet:2016}. The fact that this frequency-independent regime is not reached here seems to indicate that the acid-induced gels studied are quite dynamic, rather than completely arrested, and that this is more the case for protein gels than for droplet gels. Interestingly, this result is true over the range of volume fraction studied here, even for gels that are very concentrated.

\subsubsection{Strain dependence of the gels}

\begin{figure}[htbp]
	\centering
	\includegraphics[width=0.8\textwidth]{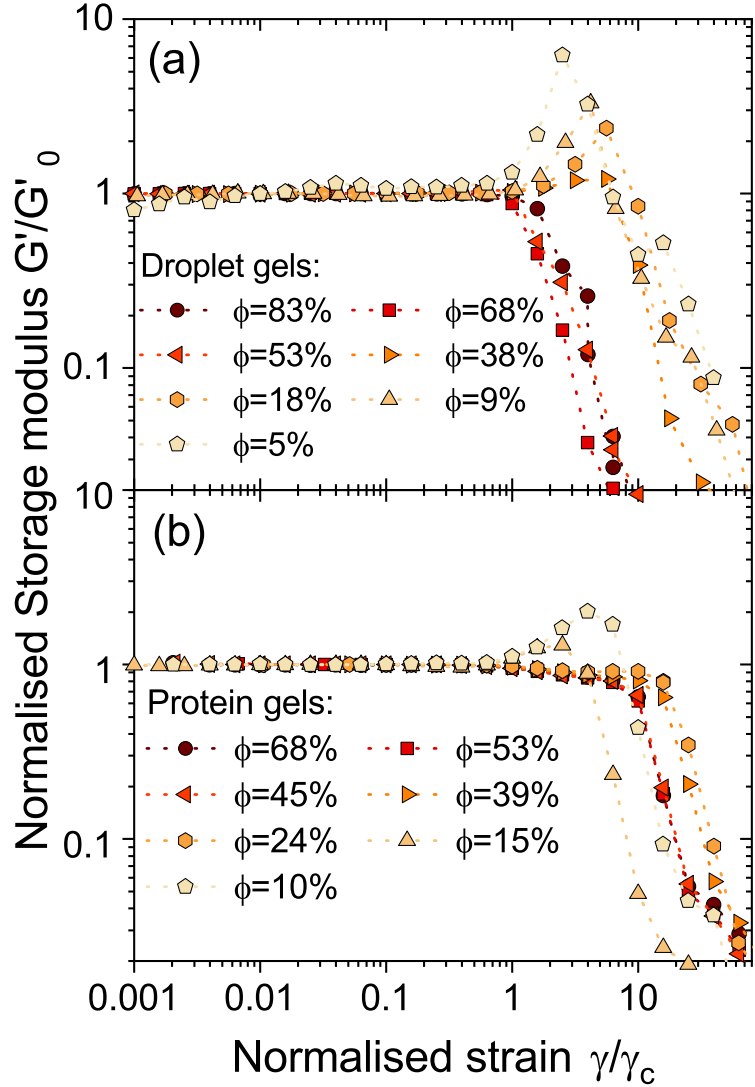}
	\caption{Storage modulus $G'$ normalised by its value in the linear regime $G'_0$ as a function of the oscillatory strain amplitude $\gamma$ normalised by its value at the onset of the non-linear regime $\gamma_c$.(a) Sodium caseinate-stabilised droplet gels at several volume fractions $\phi$ , (b) Sodium caseinate gels at several volume fractions $\phi$. 
		\\ Each curve is the average of 3 measurements, but for clarity the error bars are not represented here.}
	\label{Fig:StrainProtDrop}
\end{figure}

The oscillatory strain sweep performed on the protein and droplet gels after formation and frequency sweep, as shown in Figure~\ref{Fig:RheoMeasSeq} allows the study of the variations of the storage modulus with the amplitude of the strain oscillation. The typical strain behaviour of the gel is represented in Figure S7 , together with the definition and the values of the critical strain $\gamma_c$. To highlight the differences in strain response for all gels, this parameter was used to normalise the strain response of the gels and $G'$ was divided by its value in the linear regime. The resulting normalised curves presenting the non-linear viscoelastic behaviour pure gels of proteins and of protein-stabilised droplets at different concentrations are displayed in Figure~\ref{Fig:StrainProtDrop}.

%\textbf{Soi disant power law of critical strain vs concentration selon Wu, Morbidelli et Ruis et al. Pas du tout le cas ici, à commenter?}
%The presence of strain stiffening for multiple gels makes relevant the comparison of the different strain responses of the material with the one presented in Figure~\ref{Fig:SSProtDrop}. In order to highlight the difference,  the strain amplitude at each concentration is normalised divided by its critical value $\gamma_c$, found in Figure~\ref{Fig:StrainParam}. 

%Shall I put the error bars here? Maybe only at non-linear regime?
%Figure in supplementary material for determination of constants 

As can be seen, the nature of the non-linear regime varies with the type of gel formed and its volume fraction in proteins or droplets. The behaviour of gels at each concentration range is discussed separately below.

First, for gels prepared with suspensions of moderately low volume fraction of both proteins and protein-stabilised droplets, an increase of the normalised storage modulus $G'/G'_0$ is observed when larger shear amplitudes are applied. This phenomenon is known as strain stiffening, and this result is in correspondence with previous studies of low-concentration gels, both experimental \cite{deoliveira:2019} and computational \cite{colombo:2014b,bouzid:2018}. Using, in one case, ultrasonic imaging and, in the other, simulations of the topology of the gel networks, it was found that this behaviour could be related to irreversible stretching and reorientation of the gel branches. This behaviour was also shown to be very dependent on the structure of the network, and hence on the volume fraction of the gel. The sparser the gel, the more structural heterogeneities make possible the redistribution of the stress before failure of the material, while denser gels are more homogeneous and thus lead to a quicker breaking of bonds in the absence of reorganisation of the network. 

The strain stiffening is more pronounced, and is present on a wider range of volume fraction, for gels made of protein-stabilised droplets than for protein gels. Because strain stiffening is related to the structural heterogeneities within the network, this result may indicate that the proteins form gels that are overall more homogeneous than the droplet gels at low volume fraction, making these networks less prone to stress redistribution. The decrease in strain stiffening with the volume fraction for the two types of gels studied here is also in agreement with this phenomenological explanation. \added{As no difference in the homogeneity of the gels is visible in the micrographs in Figure~\ref{Fig:ConfocalGels}, it can only be hypothesised that the difference is at smaller length scales. }

In addition, for gels at higher concentrations, the protein gels show a slight softening in the non-linear regime, over one order of magnitude of strain amplitude, before fracture of the material. This effect is absent in the gels made of protein-stabilised droplets, where concentrated gels break at the end of the linear regime. % Can be related to another system that softens slightly?
This difference in the stress-bearing behaviour of concentrated gels may arise from structural differences between the networks; which is similar to more dilute gels. Indeed, it seems that the breakage of some bonds in the dense protein gels is not critical to the elasticity of the overall network, and leads only to a moderate decrease of $G'$ as the shear amplitude increases. On the other hand, for the dense droplet gels, the immediate drop in elasticity seems to indicate that the integrity of the whole structure is degraded upon application of a critical shear stress $\sigma_c$.

This suspected difference in the structure of the two networks would thus possibly explain the different non-linear behaviours for protein gels and droplet gels. To test this hypothesis however would require imaging each of the gel samples over a wide range of lengthscales to quantify the structural heterogeneity not only over the scale of the fractal clusters, but also over the scale of the stress-bearing backbone.

Finally, a common feature of all the protein and droplet gels is the fracture of the material at very high shear, indicated by the decrease in their elasticity. The subsequent application of low-amplitude oscilllatory shear on the gels led to no time-dependent recovery of the viscoelasticity, as presented in Figure S8, which indicates that the gel structure was irreversibly damaged. This result is in agreement with an extensive study on the fracture of caseinate gels \cite{leocmach:2014}.

\section{Conclusion}

The full sequence of rheological measurement presented in Figure~\ref{Fig:RheoMeasSeq} and confocal imaging allowed a thorough characterisation of protein gels and droplet gels by their microstructure, linear and non-linear viscoelasticity, and frequency dependence. As the two types of gels are made with colloidal particles of different nature, their behaviour was characterised over a wide range of volume fraction, in order to discriminate the intrinsic differences between the gels. Thus, in addition to the relevance of droplet gels to food products like yogurt, this comparison also yields fundamental insights into the nature of the gels. %Indeed, for the two types of gels, the inter-particle interactions arise from the destabilisation of sodium caseinate, but the particles differ in their size, nature and softness.

The first notable result is the similar properties of the two types of gels as a function of volume fraction $\phi_{eff}$, derived from the viscosity of semi-dilute suspensions \cite{roullet:2019a}. This result is significant, as the differences seen in the gel properties of proteins and protein stabilised emulsions that have been observed previously \cite{rosa:2006,chen:1999a} are accounted for by a careful choice of the composition parameter. %This result, although intuitive, is in apparent disagreement with previous studies on casein-based emulsion gels, that found emulsion gels to be markedly stronger than protein gels \cite{rosa:2006,chen:1999a}. In fact, the present study only differs by the parameter used to describe the gels, as the volume fraction was favoured over the weight concentration, and allowed a more relevant comparison. It is also interesting to note that 
The approximation of the effective volume fraction $\phi_{eff}$ held for the gels studied here, despite the complex structure of the primary colloidal particles, caseinate assemblies in one case and caseinate-coated oil droplets in the other case.
%Despite different natures of gels (dilute to quite dense, cf corresponding viscosity behaviour), uniform power-law. May hide more subtle differences.

When comparing the behaviour of protein gels and droplet gels in more detail, some differences appear between the two types of system. First, at fixed volume fraction, the droplet gels present a slightly higher elasticity than protein gels, as can be seen by the slightly higher value for the storage modulus $G'$ and the lower phase angle. Then, the viscoelasticity of protein gels is more frequency-dependent than for droplet gels, as both the storage $G'$ and the loss $G''$ moduli vary more with frequency. The  phase angle of droplet gels also displays a non-monotonic behaviour with frequency that is not seen for the protein gels. Finally, if the two types of gels at low concentrations display strain stiffening at moderate shear amplitude, this behaviour is more marked for the droplet gels, while for concentrated gels, the non-linear regime is more extended for protein gels than for droplet gels.

These minor differences seem to indicate that the theoretical framework of colloidal gels may not be sufficient for an entirely accurate description of casein gels and casein-stabilised droplet gels. It may thus be necessary to take into account some system-specific characteristics. It is possible that droplets and protein assemblies have a different inter-particle interaction, as it is believed that the single proteins adsorb at the surface of the droplets upon emulsification, and these proteins change conformation as the hydrophobic parts of their chains are anchored in the oil. Another possible explanation is that the size difference between protein assemblies and droplets leads to a different mobility of these two colloidal elements during gelation, which could be the reason for a discrepancy of microstructure and consequently of rheological behaviour. Finally, it is possible that a role is played by the softness of the particles, as the protein assemblies are soft and water-swollen, while the droplets have an incompressible oil core below the soft layer of adsorbed proteins.

%Transition to second part: Mixtures
%These characterisations of the rheological properties of pure protein and pure droplet gels can also be used to develop a better understanding of emulsion gels. In the second part of this series, mixtures of known composition are studied using the properties of the individual components to identify their contributions to the overall behaviour of the mixtures. The ability of protein-stabilised droplets to form gels of their own hints that protein-stabilised droplets in emulsion gels do not act only as fillers, as usually thought, but actively contribute to the network. 

\section{Acknowledgements}
This project forms part of the Marie Curie European Training Network COLLDENSE that has received funding from the European Union’s Horizon 2020 research and innovation programme Marie Sk\l{}odowska-Curie Actions under the grant agreement No. 642774

\appendix



%% References


\bibliographystyle{model1-num-names}
\bibliography{References}






% \end{thebibliography}

\end{document}